\documentclass[aps,pra,reprint,superscriptaddress,amsmath,amssymb]{revtex4-2}
\usepackage{graphicx}
\usepackage{xcolor}
\usepackage[colorlinks=true,bookmarks=false,citecolor=blue,urlcolor=blue]{hyperref} 

\begin{document}

\title{Spatial beam dynamics in graded-index multimode fibers under Raman amplification: a variational approach}
\author{Ashis Paul}
\affiliation{Department of Physics, Indian Institute of Technology Kharagpur, West Bengal 721302, India}
\author{Anuj P. Lara}
\affiliation{Department of Physics, Indian Institute of Technology Kharagpur, West Bengal 721302, India}
\author{Samudra Roy}
\email{samudra.roy@phy.iitkgp.ac.in}
\affiliation{Department of Physics, Indian Institute of Technology Kharagpur, West Bengal 721302, India}
\author{Govind P. Agrawal}
\affiliation{The Institute of Optics, University of Rochester, Rochester, New York 14627, USA}

\begin{abstract}
We investigate the spatial beam dynamics inside a multimode graded-index fiber under Raman amplification by adopting a semi-analytical variational approach. The variational analysis provides us with four coupled ordinary differential equations that govern the beam's dynamics under Raman gain and are much faster to solve numerically compared to the full nonlinear wave equation. Their solution also provides considerable physical insight and allows us to study the impact of important nonlinear phenomena such as self-focusing and cross-phase modulation. We first show that the variational results corroborate well with full numerical simulations and then use them to investigate the signal's dynamics under different initial conditions such as the initial widths of the pump and signal beams. This allows us to quantify the conditions under which the quality of a signal beam can improve, without collapse of the beam owing to self-focusing. While time-consuming full simulations may be needed when gain saturation and pump depletion must be included, the variational method is useful for gaining  valuable physical insight and for studying dependence of the amplified beam's width and amplitude on various physical parameters in a faster fashion.
\end{abstract}
\maketitle

\section{Introduction}

The quest for higher and higher output powers has led to the use of multimode fibers for fiber-based lasers and amplifiers \cite{jauregui2013high, zervas2014high, zuo2022high}.
In recent years, graded-index (GRIN) multimode fibers have replaced traditional step-index fibers for making high-power Raman amplifiers. This is motivated by a phenomenon known as the Raman-induced spatial beam cleanup \cite{baek_single-mode_2004, terry_explanation_2007, terry_beam_2008, kuznetsov_beam_2020, Sidelnikov2022}, which improves considerably the amplified signal's beam quality at the output end of a Raman amplifier. Recent experiments have shown that power levels of more than 2~kW can be realized using GRIN fibers for Raman amplification \cite{chen2020,chen2021,Fan2021}.

The use of a mode-based approach for understanding the Raman-induced spatial beam cleanup
becomes less appropriate when many modes of a GRIN fiber are excited by the incoming pump and signal beams. A non-modal numerical approach has recently been proposed for both Yb-doped fiber amplifiers \cite{jima_numerical_2022} and Raman GRIN amplifiers\cite{Sidelnikov2022} that takes into account most relevant physical effects under continuous-wave (CW) diode pumping. However, such an extensive numerical model is time-consuming because it requires a solution of the coupled nonlinear partial differential equations satisfied by the pump and signal beams. Recent work on Kerr-induced beam cleaning \cite{liu_kerr_2016, krupa_spatial_2017} has shown that the phenomenon of periodic self-imaging \cite{agrawal_invite_2019}, a unique property of GRIN fibers, plays an important role in the amplification of the signal beam by creating a nonlinear index grating inside the GRIN fiber. Simple analytic models based on self-imaging have been recently proposed to study amplification and beam narrowing in GRIN fibers \cite{agrawal_spatial_2023, agrawal_spatial_2023-1}. However, these models did not include fully the important nonlinear effects such as self-phase modulation (SPM) and cross-phase modulation (XPM).

In this work, we apply the variational method to develop a semi-analytic model of the Raman-amplification process in GRIN fiber amplifiers. We solve the resulting equations numerically to investigate the impact of both SPM and XPM on the performance of a Raman amplifier. In our treatment, the Gaussian profile of the pump is not approximated with a parabolic shape, which may produce erroneous results. We show that our semi-analytic approach is much less time-consuming computationally compared to a fully numerical approach and that it also provides considerable physical insight. The article is organized as follows. In Section II, we outline the theory that leads to the nonlinear coupled propagation equations for the pump and signal beams. In section III, we develop the variational analysis by forming a suitable Lagrangian and derive four coupled ordinary differential equation describing the signal beam's dynamics under the Raman gain. We compare the variational results with full numerical simulations and show the robustness of our approach. In Section IV, we solve numerically the coupled equations  for the signal beam's parameters and investigate its dynamics under the impact of SPM and XPM for different initial conditions. The main conclusions are summarized in Section V.

\section{Theory}

\begin{figure}
	\includegraphics[width=\linewidth ]{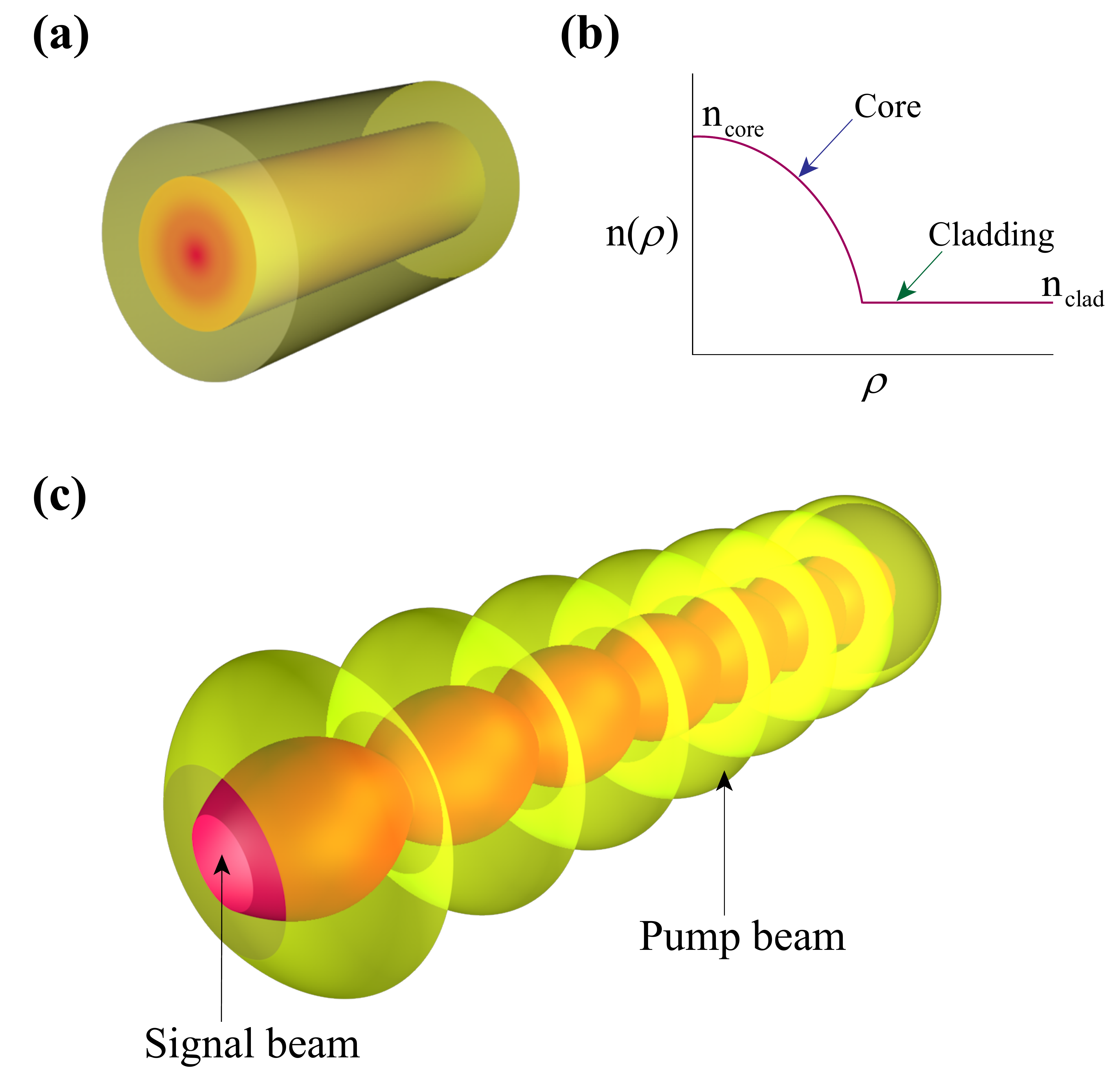}
\caption{\textbf{(a)} Schematic of a GRIN fiber and \textbf{(b)} its parabolic index profile. \textbf{(c)} Two isosurfaces showing how the signal and pump beams evolve in a periodic fashion inside a GRIN fiber because of the self-imaging provided by the parabolic index profile.} \label{Fig1}
\end{figure}

We consider a GRIN fiber with a parabolic refractive index  profile (see parts (a) and(b) of Fig.~\ref{Fig1}) and include the optical Kerr effect using
\begin{equation}
    n(\rho) = n_{\rm core} \left(1-\frac{1}{2}b^2 \rho^2\right) + n_2 |E|^2,
    \label{Eq1}
\end{equation}
where $\rho = \sqrt{x^2+y^2}$ is the radial distance from the central axis of the GRIN fiber and $n_{\rm core}$ is the refractive index at $\rho =0$. The index gradient $b$ is defined as $b = \sqrt{2\Delta}/a$, where $a$ is the core's radius of the GRIN fiber and
$\Delta$ is the relative core--cladding index difference defined as, $\Delta=(n_{\rm core}-n_{\rm clad})/n_{\rm core}$. The Kerr coefficient $n_2$ and has a value of $2.7 \times 10^{-20} \rm m^2/W$ for silica fibers. In practice, silica molecules also produce a significant nonlinear response, in addition to the instantaneous Kerr response of electrons. This response, known as the Raman response, is delayed in time. It is accounted for by modifying the Kerr term $n_2 |E|^2$ in Eq. (\ref{Eq1}) as~\cite{GPAbook}
\begin{equation}
    n_2 |E|^2 = (1-f_{R}) n_2 |E|^2 + f_{R} n_2\int_{0}^{\infty}h_{R}(t')|E(t-t')|^2 dt',
    \label{eq2}
\end{equation}
where $f_{R}$ is the fractional Raman contribution (about 18\% for silica fibers) and $h_{R}(t)$ is the Raman response function, normalized with $\int_0^{\infty} h_{R}(t)dt = 1$.

The pump and signal beams are launched at the input end of the GRIN fiber located at $z=0$. The total electric field $E(\mathbf{r},t)$ inside the fiber at a distance $z$ can be written as
\begin{equation}
   E (\mathbf{r},t) = A_p {\rm exp} \left[i\left(k_p z - \omega_p t \right) \right]
   + A_s {\rm exp} \left[i\left(k_s z - \omega_s t \right) \right],
   \label{eq3}
\end{equation}
where $A_j$ and $k_j = n_{core}(\omega_j) \omega_j/c$ with $j = p, s$ are  the amplitudes and wave numbers of pump and signal beams, respectively. Both waves are assumed to be polarized along the same direction. In our simulations, we choose the wavelengths for the pump ($\lambda_p$) and signal ($\lambda_s$) beams to be 1018~nm and 1060~nm, respectively and use $\Omega = \omega_p - \omega_s$ for the frequency shift of the signal from the pump. Using Eq.\ (\ref{eq3}), we evaluate the integral in Eq.~(\ref{eq2}) to obtain
\begin{align}
    \int_{0}^{\infty} h_{R}(t') |E(t-t')|^2 dt' = |A_p|^2 + |A_s|^2 \nonumber \\
    + A_p^* A_s e^{-i\delta k z} \tilde{h}_R(\Omega) + A_p A_s^* e^{i\delta k z} \tilde{h}_R^*(\Omega),
    \label{eq4}
\end{align}
where $\delta k = k_p - k_s$ and $\tilde{h}_R (\Omega)$ is the Fourier transform of $h_R(t)$. The Raman gain coefficient $g_R$ is related to the imaginary part of $\tilde{h}_R (\Omega)$ as $g_R = 2 f_R n_2 (\omega_s/c) {\rm Im} (\tilde{h}_R)$.

Using Eqs.\ (\ref{eq2})-(\ref{eq4}) in Maxwell's equations and retaining only the phase-matched terms under the slowly varying envelope approximation, we can separate the pump and signal terms and obtain the following two coupled nonlinear equations for the pump and signal amplitudes~\cite{agrawal_spatial_2023},
\begin{align}
	\frac{\partial A_p}{\partial z} + \frac{\nabla_{\bot}^2 A_p}{2ik_p} + \frac{i}{2}k_p b^2 \rho^2 A_p &= \frac{i\omega_p}{c} n_2 |A_p|^2 A_p \label{eq7}, \\
	\frac{\partial A_s}{\partial z} + \frac{\nabla_{\bot}^2 A_s}{2ik_s} + \frac{i}{2}k_s b^2 \rho^2 A_s &= \frac{i\omega_s}{c} n_2 \left(|A_s|^2 + 2|A_p|^2 \right) A_s\notag\\  
    &+ \frac{1}{2} g_R |A_p|^2 A_s, \label{eq8}
\end{align}
where $\nabla_{\bot}^2 = \partial^2/\partial x^2 + \partial^2/\partial y^2$ is the transverse Laplacian. Here, we have neglected the depletion and XPM terms in the pump equation [Eq.~\ref{eq7}], assuming the pump to remain much more intense than the signal over entire length of the GRIN fiber. As a result, this equation can be solved first to obtain $A_p (\rho, z)$. For a CW pump in the form of a Gaussian beam, Eq.\ (\ref{eq7}) has been solved with the variational method and the solution is given as~\cite{karlsson_dynamics_1992}
\begin{equation}
    A_p(\rho,z) = \sqrt{\frac{I_{p0}}{f_p(z)}} {\rm exp} \left[-\frac{\rho^2}{2 w_{p0}^2 f_p(z)} + i \Theta(\rho,z)\right],
    \label{eq9}
\end{equation}
where $I_{p0}$ is the peak intensity and $w_{p0}$ is the width of the input pump beam
at $z=0$, while the periodic function $f_p(z)$ is defined as
\begin{equation}
    f_p (z) = \cos^2(bz) + C_p^2 \sin^2(bz),~~C_p = \sqrt{1-p}/(bk_p w_{p0}^2),
    \label{eq9b}
\end{equation}
with $p = P_{p0}/P_c$. Here $P_{p0} = (\pi w_{p0}^2) I_{p0}$ is the input pump power and $P_c = 2\pi n_{core }/(n_2 k_p^2)$ is the critical  power at which the pump beam collapses due to self-focusing. Because of GRIN-induced self-imaging, the pump beam compresses and expands periodically such that it recovers its input shape and width at distances  $z = m \pi/b$, where $m$ is an integer. Figure \ref{Fig1}(c) shows schematically the periodic evolution of such a pump beam inside the GRIN fiber.

\section{Variational analysis}

Exploiting the analytical pump solution given in Eq. (\ref{eq9}), we can  numerically solve the signal equation in Eq.\ (\ref{eq8}). However, numerical simulations become time-consuming for distances exceeding $>$10~m required for Raman amplifiers. A numerical approach also hinders physical insight and does not reveal what parameters are most relevant for narrowing of the signal beam to occur. To gain insight into how the the SPM, XPM, and Raman phenomena affect the signal beam, we adopt  the variational method \cite{anderson_variational_1983} for solving  Eq.~(\ref{eq8}). The variational method has been used successfully, in spite of the gain and loss terms that make the underlying system non-conservative \cite{Roy2009}. It requires a suitable ansatz for the pulse shape and makes the assumption that the functional form of the pulse shape remains intact in the presence of small perturbations, but its parameters appearing in the ansatz (amplitude, width, position, phase, frequency, etc.) evolve
with propagation.

First, we normalize Eq.~(\ref{eq8}) and rewrite it in the following form,
\begin{align}
    i\frac{\partial \psi_s}{\partial \xi} &+ \frac{\delta}{2}\left(\frac{\partial^2 \psi_s}{\partial r^2} + \frac{1}{r} \frac{\partial \psi_s}{\partial r} \right) - \frac{1}{2\delta} r^2 \psi_s  + \gamma |\psi_s|^2 \psi_s = \notag \\
    &-2 \Gamma |\psi_p|^2 \psi_s + i\frac{G_R}{2}|\psi_p|^2 \psi_s, \label{eq10}
\end{align}
where the variables were scaled using $\xi= bz$, $r=\rho/w_{s0}$, and $\psi_j= A_j/\sqrt{I_{j0}}$ with $j = s,p$. Here $w_{s0}$ is the width and $I_{s0}$ is the peak intensity of the input signal beam, $\delta = w_g^2/w_{s0}^2 $ is a dimensionless ratio, and $w_g^2 = 1/(bk_s)$ is the width of the fundamental mode of the GRIN fiber at the signal's frequency $\omega_s$. Typically $w_g$ is close to 5~$\mu$m for GRIN fibers. Other parameters, $\gamma = \omega_s n_2 I_{s0}/(cb)$,  $\Gamma = \omega_s n_2 I_{p0}/(cb)$, and $G_R = g_R I_{p0}/b$, are the normalized SPM, XPM, and Raman coefficients respectively. To implement the variational method, we treat the two terms on the right side of Eq.~(\ref{eq10}) as a small perturbation $\epsilon$, defined such that
\begin{align}
    \epsilon = \left[2i\Gamma + (1/2) G_R\right] |\psi_p|^2 \psi_s.
    \label{eq11}
\end{align}

The Lagrangian density $\mathcal{L}_d$ corresponding to Eq.~(\ref{eq10}) has the form \cite{anderson_pereira-stenflo_1999}
\begin{align}
    \mathcal{L}_d =\frac{i}{2} r &\left(\psi_s \partial_{\xi} \psi_{s}^* - \psi_{s}^* \partial_{\xi} \psi_{s} \right) + \frac{\delta}{2} r |\partial_r \psi_s|^2 \nonumber\\ 
    &- \frac{\gamma}{2} r |\psi_s|^4  + \frac{r^3}{2\delta} |\psi_s|^2 + i r \left(\epsilon \psi_s^* - \epsilon^* \psi_s\right), \label{eq12}
\end{align}
where $\partial_{j} \equiv  \partial/\partial j$.  We choose a chirped Gaussian beam for our ansatz for $\psi_s$ because the signal is often in the form of a Gaussian beam in practice:
\begin{equation}
    \psi_s(r,\xi) = \psi_{s0}(\xi) \; {\rm exp} \left[-\frac{r^2}{r_s^2(\xi)}  + i d_s(\xi) r^2 + i \phi_s(\xi) \right], \label{eq13}
\end{equation}
where the four parameters, $\psi_{s0}, r_s , d_s$ , and $\phi_s$ depend on $\xi$. Using this ansatz and following the standard Rayleigh--Ritz optimization procedure \cite{anderson_pereira-stenflo_1999}, we obtain the reduced Lagrangian, $L = \int_0^\infty \mathcal{L}_d dr$, by integrating over $r$. The result is found to be
\begin{align}
    L &=& \frac{1}{2} \psi_{s0}^2 r_s^2 \left(\frac{d\phi_s}{d\xi}\right) + \left[2 \delta d_s^2 + \frac{1}{2\delta} + \frac{d d_s}{d\xi}\right] \frac{\psi_{s0}^2 r_s^4}{2} \nonumber \\ && + \frac{\delta}{4} \psi_{s0}^2 - \frac{\gamma}{8} \psi_{s0}^4 r_s^2 + i \int_0^\infty r \left(\epsilon \psi_s^* - \epsilon^* \psi_s \right) dr.
\label{eq14}
\end{align}
We evaluate the integral in Eq.~(\ref{eq14}) by using the pump intensity from Eq.~(\ref{eq9}). The normalized form of this intensity is
\begin{equation}
    |\psi_p (\xi)|^2 = f_p(\xi)^{-1} {\rm exp}\left[{-r^2/\left(r_{p0}^2 f_p(\xi)\right)}\right]. \label{eq14b}
\end{equation}

We now use the \textit{Euler-Lagrange} equation, $\partial_\xi (\partial_{X_\xi} L) - \partial_X L = 0$, with $X = \psi_{s0}, r_s, d_s, \phi_s $ and obtain the following four coupled equations for the evolution of the four parameters along the amplifier's length:
\begin{align}
    \frac{d \psi_{s0}}{d \xi} &= - 2 \delta d_s \psi_{s0} + \frac{G_R}{2f_p} \left(\frac{r_e}{r_s}\right)^2 \left[2-\left(\frac{r_e}{r_s}\right)^2\right] \psi_{s0},
    \label{eq15} \\
    \frac{d r_s}{d \xi} &= 2 \delta d_s r_s - \frac{G_R}{2f_p} \left(\frac{r_e}{r_s}\right)^2 \left[1-\left(\frac{r_e}{r_s}\right)^2\right] r_s, \label{eq16} \\
    \frac{d d_s}{d\xi} &= - 2 \delta d_s^2 - \frac{\gamma}{4} \left( \frac{\psi_{s0}}{r_s} \right)^2 - \frac{1}{2\delta} + \frac{\delta}{2 r_s^4} \nonumber \\
    &- \frac{2\Gamma}{f_p} \left(\frac{r_e^2}{r_s^4}\right) \left[1-\left(\frac{r_e}{r_s}\right)^2\right],\label{eq17} \\
    \frac{d\phi_s}{d\xi} &= -\frac{\delta}{r_s^2} + \frac{3}{4} \gamma\psi_{s0}^2 + \frac{2 \Gamma}{f_p} \left(\frac{r_e}{r_s}\right)^2 \left[2-\left(\frac{r_e}{r_s}\right)^2\right],    \label{eq18}
\end{align}
where $r_e^{-2}= r_s^{-2}+r_{p0}^{-2}/ f_p$ with $f_p = \cos^2(\xi) + C_p^2 \sin^2(\xi)$. These ordinary differential equations (ODEs) can be solved numerically much faster than Eq.~(\ref{eq10}) governing the evolution of the signal beam. However, accuracy of the resulting solution needs to be checked by solving Eq.~(\ref{eq10}) directly.

\begin{figure} [tb!]
	\includegraphics[width=\linewidth]{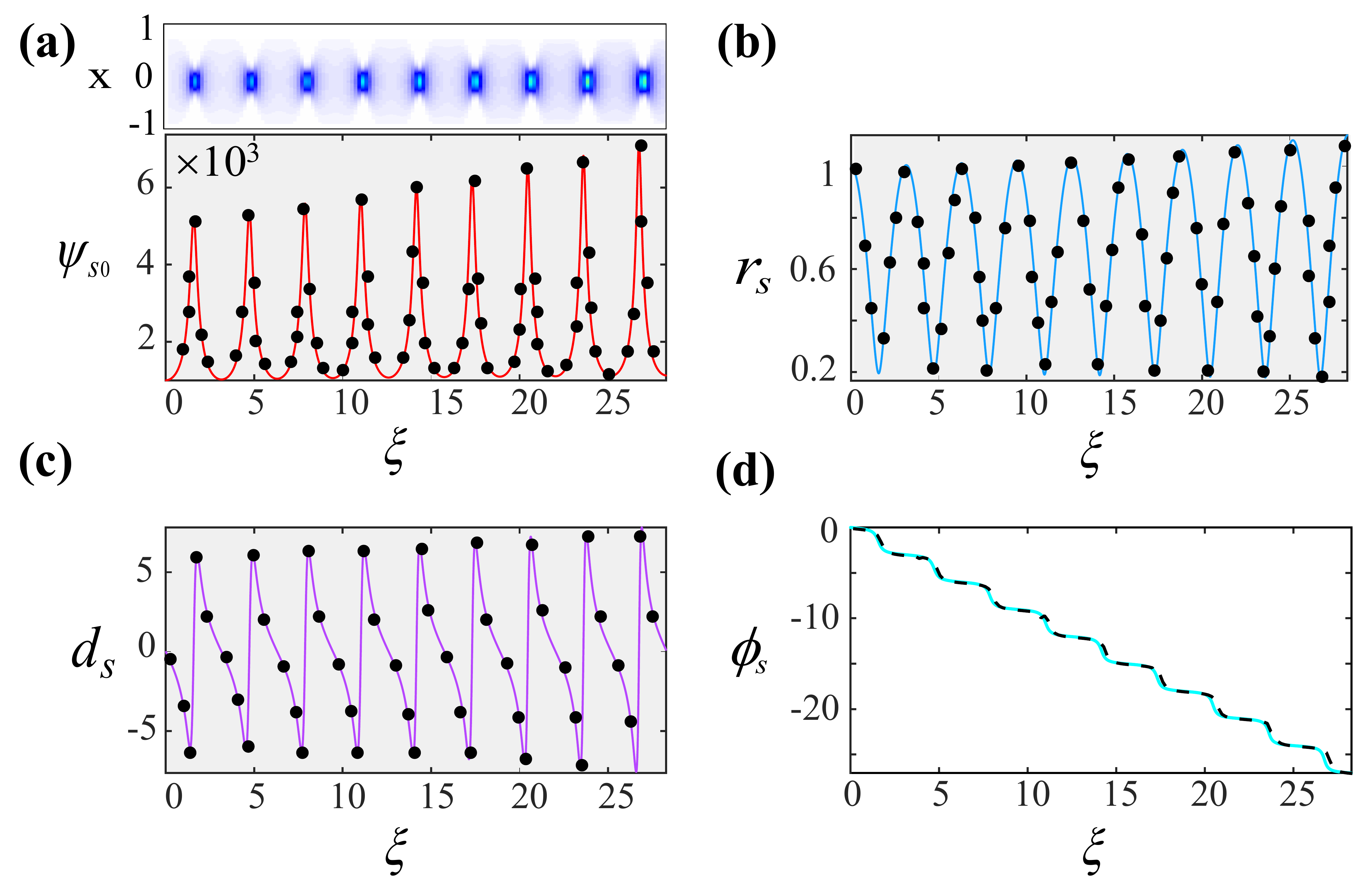}
\caption{Comparison between the variational (solid lines) and numerical (solid dots and dashes) predictions: \textbf{(a)} amplitude, \textbf{(b)} width, \textbf{(c)} phase-front curvature, and \textbf{(d)} phase of the signal beam. The top inset in \textbf{(a)} shows periodic self-imaging of the signal beam inside a GRIN Raman amplifier (see text for simulation parameters).}
	\label{Fig2}
\end{figure}

We solve the coupled ODEs, Eqs.\ (\ref{eq15})-(\ref{eq18}), with the fourth-order Runge--Kutta method and obtain the evolution of four beam parameters under different conditions. We check accuracy of the solution by solving Eq.~(\ref{eq10}) numerically with the standard \textit{split-step Fourier} (SSF) method. In both cases, we employ the same values of the three parameters: $\gamma=1 \times 10^{-3}$, $\Gamma=5.7 \times 10^{-3}$ and $G_R=3.5 \times 10^{-3}$, which were estimated using the following realistic values for a GRIN fiber: $a=50$ $\mu$m and $\Delta=0.1$, making $b=2.83 \times 10^{3}$ m$^{-1}$. The input pump power is taken to be $P_{p0}= 0.1$~MW, and signal power $P_{s0}$ is 1$\%$ of $P_{p0}$. The input beam has a Gaussian shape such that $\psi_s(r)=\psi_{s0}\exp(-r^2/r_s^2 + id_s r^2)$ with the initial values $\psi_{s0}=1$, $r_{s}=1$, and $d_{s}=0$. Figure \ref{Fig2} compares the variational and numerical results over a propagation distance that corresponds to nine self-imaging periods ($\xi \approx 28$). An excellent agreement between the numerical and variational results is evident in Fig.~\ref{Fig2} over this distance.

\section{Impact of SPM and XPM}

Since the variational method predicts accurately the signal beam's dynamics under Rman amplification, we use it to study two different scenarios for the signal's input width: (i) signal beam is narrower than the pump beam ($w_{s0}< w_{p0}$) and (ii) signal beam is wider than the pump beam ($w_{s0}> w_{p0}$). We use realistic values for all other parameters.

\subsection{Case I:  $w_{s0}< w_{p0}$}

Assuming $w_{p0} = 1.33 \times w_{s0}$, we solve Eqs.\ (\ref{eq15})--(\ref{eq18}) with and without including the SPM and XPM effects. In the absence of XPM and SPM, we set $\Gamma = 0$ and $\gamma = 0$. This case was studied recently \cite{agrawal_spatial_2023}, and it was found that, in the absence of the Raman gain, the width satisfies a simple equation,
\begin{equation}
    \frac{d^2 r_s}{d\xi^2} +r_s=\delta^2r_s^{-3},
\end{equation}
and has the analytic solution $r_s(\xi)=[\cos^2(\xi)+ \delta^2\sin^2(\xi)]^{1/2}$.  Following \cite{agrawal_spatial_2023}, we define the dimensionless parameter,  $\mu = g_R I_{p0} w_{s0}^2 / (2bw_{p0}^2)$, which is controlled mainly by the input pump intensity (or power). For silica-based GRIN fibers, the Raman gain coefficient is $g_R = 1 \times 10^{-13}$ m/W at wavelengths near $1~\mu$m. At an input pump power of $115$ kW, the peak intensity is close to $100$ TW/m$^2$. Its use yields $\mu = 0.001$ and $C_p = 0.1$ for $r_{p0}/r_{s0} = 1.33$. Such high power values can be realized when the pump is in the form of nanosecond pulses.

\begin{figure}[tb!]
	\includegraphics[width=\linewidth ]{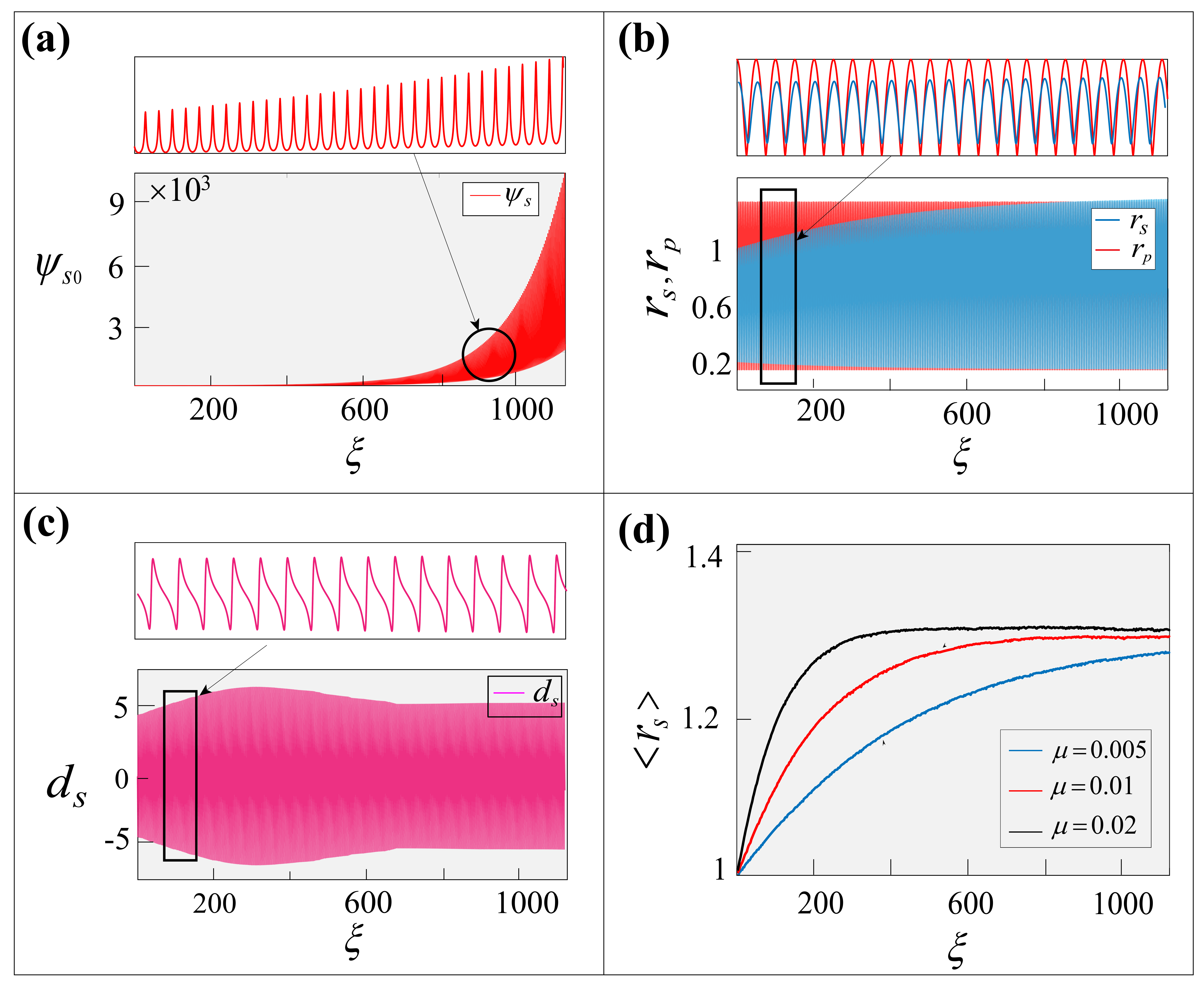}
\caption{\textbf{(a)-(c)} Evolution of three parameters of the signal beam ($\psi_{s0}, r_s, d_s$) without the XPM and SPM effects ($\gamma=\Gamma=0$) using $r_{p0}= 1.33r_{s0}$ for a distance $\xi=1100 ~(z\approx 40$ cm). The insets show the oscillatory pattern on a magnified scale. \textbf{(d)} Evolution of average signal width $\left \langle r_s\right\rangle$  for three values of $\mu$ showing saturation of the width for large $\xi$ values. }
	\label{Fig3}
\end{figure}

Setting $\Gamma =0 $, $\gamma =0$, $\delta = 0.2$, and $\mu = 0.001$, we solved Eqs.\ (\ref{eq15})-(\ref{eq18}), and the results are shown in Fig.~\ref{Fig3}. As expected, the Raman gain amplifies the signal, but its amplitude in Fig.~\ref{Fig3}(a) exhibits an oscillatory pattern, seen clearly in the magnified view on top. In Fig.~\ref{Fig3}(b), the signal's width also follows the same periodic pattern, as dictated by periodic self-imaging of the pump width $r_p$ shown in red. In the zoomed-in version on top, at the point of maximum compression, where the effect of pump on the signal is most prominent, signal's width is larger than the pump's width. To visualize changes in the signal's width more clearly, we plot in part (d) of Fig.~\ref{Fig3} its average value $\langle r_s \rangle$ as a function of propagation distance $\xi$ for three $\mu$ values. The average width was calculated by averaging $r_s$ over several self-imaging periods $\xi_p$ and then dividing it by the initial average width to normalize. Clearly, the average signal width gradually increases and saturates at a value close to the pump's width for large $\xi$. The rate of increase depends on the value of $\mu$ and is larger for its larger values.

\begin{figure}[tb!]
	\includegraphics[width=\linewidth ]{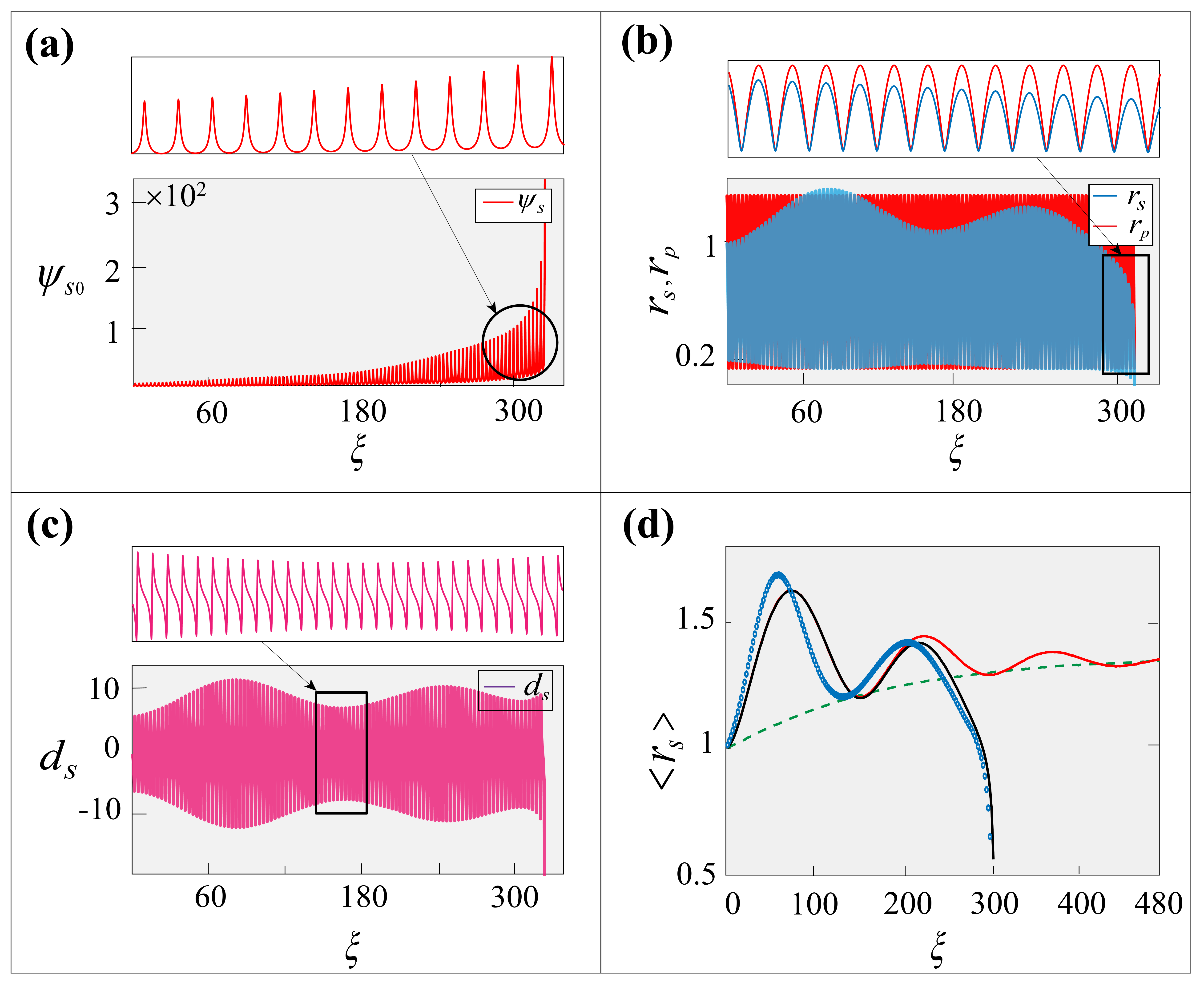}
\caption{\textbf{(a)-(c)} Same as Fig.~\ref{Fig3} but the nonlinear SPM and XPM effects are included using $\gamma = 1\times 10^{-3}$ and $\Gamma = 6\times 10^{-3}$. \textbf{(d)}  Impact of the nonlinear effects on of the average signal's width $\langle r_s \rangle$: evolution  without SPM and XPM (green dotted), with XPM only (red solid), and with both SPM and XPM (black solid). In the last case, numerically predicted values are shown by blue circles.}
	\label{Fig4}
\end{figure}

The important question is how the behavior seen in Fig.~\ref{Fig3} is modified when the nonlinear effects (both SPM and XPM) are included. This question is answered by the results shown in Fig.~\ref{Fig4}, where we included the nonlinear effects by using $\gamma = 1 \times 10^{-3}$ and $\Gamma = 6 \times 10^{-3}$ but kept all other parameters the same. While all three parameters oscillate during propagation in parts (a)-(c), an important difference is that the signal beam collapses at a certain distance in the presence of SPM such that its width approaches zero [see Fig. \ref{Fig4}(c)]. The collapse is seen more clearly in part (d), where $\langle r_s \rangle $ is plotted as a function of $\xi$ under three different conditions. When both SPM and XPM are neglected, $\langle r_s \rangle $ increases monotonically towards a saturated value, as shown by the green dotted line. When only XPM is included (red solid line), $\langle r_s \rangle $ exhibits damped oscillation and its value saturates at the pump's width, without any beam collapse. When both XPM and SPM are included, $\langle r_s \rangle $ oscillates initially but eventually collapses (black solid line) toward the value zero as the amplification of the signal beam increases the beam's power toward the critical self-focusing power. It may appear surprising that a variational analysis does not break down near the self-focusing collapse. To verify its accuracy, we solved the full wave equation, Eq.~(\ref{eq10}) and extracted numerically the values of $\langle r_s \rangle $.  The results are indicated in Fig.~\ref{Fig4}(d) by blue circles, and they show satisfactory agreement with the variational results.

\subsection{Case II: $w_{s0}> w_{p0}$}

In this subsection, we consider the evolution of a signal beam when its initial width is greater than pump's width. For numerical simulations, we choose $r_{p0}/r_{s0}=0.67$ with $\mu=0.01$, $C_p=0.42$ and $\delta= 0.2$. As before, we solve Eqs.\ (\ref{eq15})-(\ref{eq18}) with and without the nonlinear SPM and XPM terms and study how the signal beam's parameters change along the Raman amplifier's length.

\begin{figure}[tb!]
	\includegraphics[width=\linewidth ]{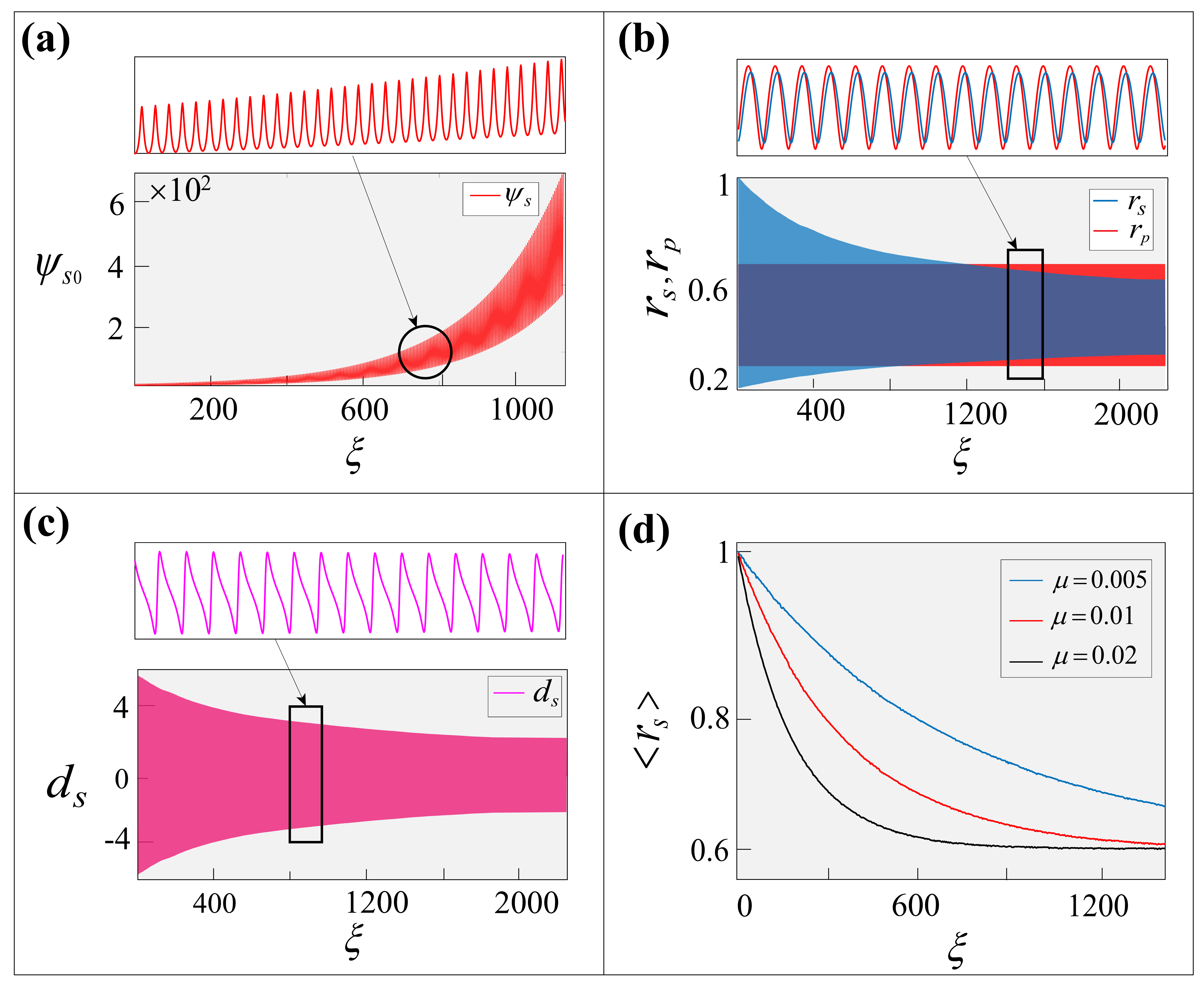}
\caption{Variational solution for the signal's parameters without the XPM and SPM effects ($\gamma=\Gamma=0$) when $w_{s0}>w_{p0}$. The results are shown in the same format used for Fig.~\ref{Fig3}. Parts \textbf{(a)-(c)} show the evolution of three parameters (amplitude, width, and chirp). Part \textbf{(d)} shows how the average width $ \langle r_s \rangle$ changes with $\xi$ for different values of $\mu $.}
	\label{Fig5}
\end{figure}

In the absence of both XPM and SPM $(\Gamma = 0, \gamma = 0)$, the results are shown in Fig.~\ref{Fig5} in the same format used for Fig.~\ref{Fig3}. We used $P_{p0}=75$ kW for the pump's power and set $P_{s0}$ at 1\% of $P_{p0}$, resulting in $G_R=9\times 10^{-3}$. The evolution of signal's width, shown in  Fig.~\ref{Fig5}(b), shows beam narrowing similar to that reported in Ref.~\cite{agrawal_spatial_2023}. The narrowing of signal beam is seen more clearly in part (d) of Fig.~\ref{Fig5}, where we plot $ \langle r_s \rangle $ as a function of distance $\xi$ for different $\mu$ values. The average width of the signal decreases in an exponential fashion and reaches a steady-state value close to the pump's initial width at large distances. For higher values of $\mu$, saturation of the beam-width occurs at shorter propagation distances.

\begin{figure}[tb!]
	\includegraphics[width=\linewidth ]{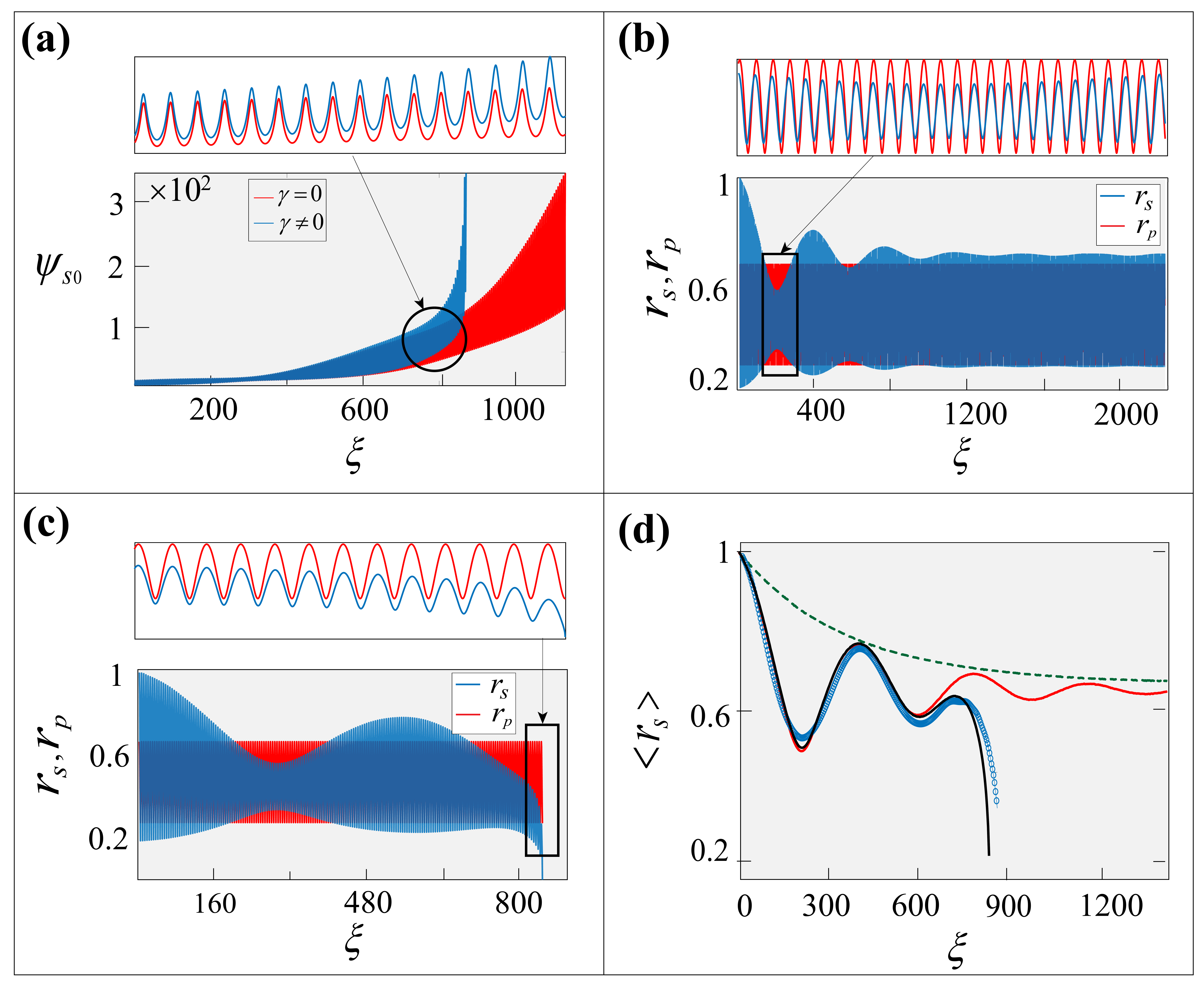}
	\caption{Impact of the XPM and SPM effects on the signal's parameters in the case $w_{s0}>w_{p0}$. \textbf{(a)} Evolution of signal's amplitude with (blue) and without (red) the SPM term. \textbf{(b)} Evolution of signal's width with XPM but without SPM ($\gamma=0$).  \textbf{(c)} Evolution signal's width when both SPM and XPM are included. Note the beam collapse occurring in this situation. \textbf{(d)} Comparison of changes in the average width $\langle r_s \rangle $ in different regimes: without SPM and XPM (green dotted), with XPM only (red solid), both SPM and XPM included (black solid). Numerical prediction in the last case are shown by blue circles.}
	\label{Fig6}
\end{figure}

Again, the question is how the behavior seen in Fig.~\ref{Fig5} is modified when the nonlinear effects (both SPM and XPM) are included. This question is answered by the results shown in Fig.~\ref{Fig6}, where we include the nonlinear effects by using finite values of $\gamma$ and $\Gamma$ while keeping all other parameters the same. The signal's evolution changes significantly under the influence of SPM and XPM\@. In part (a), we compare the growth of signal's amplitude in the presence (blue line) and absence (red line) of SPM, while keeping the XPM term in both cases using $\Gamma=1.4 \times 10^{-2}$. For the blue curve $\gamma$ was $6.45 \times 10^{-4}$. Changes in the signal and pump width are shown in part (b) with XPM included without SPM ($\gamma=0$). Both widths follow an oscillatory pattern indicative of self-imaging inside a GRIN fiber. Part (c) shows the same two widths after including the SPM effects as well. It is evident that the width evolutions changes completely when SPM is included. This is so because the beam collapses occurs when the signal power approaches the critical power required for self-focusing. Plot (d) compares the evolution of average beam width $\langle r_s \rangle$ in three cases. This width decays monotonically when SPM and XPM effects are weak or ignored, exhibits damped oscillations when XPM is included without SPM, and experiences self-focusing collapse when both  XPM  and SPM are included. Narrowing of the signal beam with increasing amplification expedites the self-focusing effect.

\section{Conclusion}

We have studied the Raman amplification of a signal beam inside a multimode graded-index (GRIN) fiber with a semi-analytical variational approach, assuming that both the pump and signal are launched into the GRIN fiber in the form of CW or quasi-CW Gaussian beams. The variational analysis provides us with four coupled ordinary differential equations for the four relevant parameters (amplitude, width, phase, and phase curvature) that govern the evolution of the signal beam inside a GRIN fiber. These equations are much faster to solve numerically compared to the coupled nonlinear wave equation satisfied by the pump and signal beams. Their solution also provides considerable physical insight and allows us to study the impact of important nonlinear phenomena such as SPM, XPM, and self-focusing.

We first verify the accuracy of the variational equations by comparing their solution to the numerical prediction of the nonlinear wave equation governing the signal-beam's evolution when the pump's depletion is ignored. We then use the variational equations for investigating the signal's evolution inside the GRIN fiber under different initial conditions such as the initial widths of the pump and signal beams. This allows us to quantify the conditions under which the quality of a signal beam can improve, without its collapse owing to self-focusing. Two conclusions can be drawn from our results given in this paper. First, narrowing of the signal beam can occur only when the input width of the pump beam is comparable or smaller than that of the signal. As a result, Raman-induced beam cleanup is unlikely to occur in cladding pumped Raman amplifiers where pump beam is always considerably wider than the signal beam.
Second, one must avoid the collapse of the signal beam induced by self-focusing. In practice, this can be realized by decreasing the input signal power or the length of the GRIN fiber to ensure that the signal's power remains below the critical power at which self-focusing leads to beam's collapse.

While time-consuming full simulations may be needed when gain saturation and pump depletion must be included, the variational method is useful for gaining valuable physical insight and for studying in a much faster fashion dependence of the amplified beam's width and amplitude on various physical parameters. We have shown that the variational result agree well with full numerical simulations and provide significant physical insights.

\section{Acknowledgement}
A.P. acknowledges the Ministry of Human
Resource Development, Government of India and IIT
Kharagpur for financial support to carry out his research
work. A.P.L. acknowledges University Grants Commission, India
for support through Junior Research Fellowship in Sciences,
Humanities and Social Sciences (ID 515364).

\bibliographystyle{apsrev4-2}
\bibliography{reference}

\end{document}